\documentstyle[12pt,epsf]{article}
 \catcode`@=11
\long\def\@caption#1[#2]#3{\par\addcontentsline{\csname
  ext@#1\endcsname}{#1}{\protect\numberline{\csname
  the#1\endcsname}{\ignorespaces #2}}\begingroup
    \small
    \@parboxrestore
    \@makecaption{\csname fnum@#1\endcsname}{\ignorespaces #3}\par
  \endgroup}
\catcode`@=12
\newcommand{\bi}{\bibitem}
\begin{document}
\setlength{\baselineskip}{0.27in}

\newcommand{\beq}{\begin{equation}}
\newcommand{\eeq}{\end{equation}}
\newcommand{\beqa}{\begin{eqnarray}}
\newcommand{\eeqa}{\end{eqnarray}}
\newcommand{\lsim}{\begin{array}{c}\,\sim\vspace{-21pt}\\<
\end{array}}
\newcommand{\gsim}{\begin{array}{c}\sim\vspace{-21pt}\\>
\end{array}}
\def\OMIT#1{}
\newcommand{\bfsigma}{\boldmath \sigma\unboldmath}
\newcommand{\bfpartial}{\boldmath \partial\unboldmath}

\begin{titlepage}
{\hbox to\hsize {\hfill UCSD-97-06}}

\begin{center}
\vglue .06in
{\Large \bf Effective Field Theory and 
Matching in Non-Relativistic Gauge Theories}
\\[.5in]

\begin{tabular}{c}
{\bf Benjamin Grinstein and  Ira Z. Rothstein}\\[.05in]
{\it Department of Physics}\\
{\it University of California at San Diego}\\
{\it La Jolla, CA 92122 }\\[.15in]
\end{tabular}
\vskip 0.25cm
\vskip 0.25cm

{\bf Abstract}\\[-0.05in]

\begin{quote}
The effective  Lagrangian and power counting rules 
for non-relativistic gauge theories 
are derived via a  systematic expansion in the large $c$ limit.
It is shown that the $1/c$ expansion leads to an effective
field theory which incorporates a multipole expansion.
Within this theory  there is no need for heuristic
arguments to determine the scalings of operators. After eliminating
$c$ from the lowest order Lagrangian the states of the theory become
independent of $c$ and the scaling of an operator is given simply by
its overall coefficient. 
We show how this power counting works in the calculation of the
Lamb shift within the effective field theory formalism.

\end{quote}
\end{center}
\end{titlepage}
\newpage

Effective field theories are indispensable tools for studying systems
with disparate scales. The idea dates back to the Euler-Heisenberg
Lagrangian for QED \cite{IZ} and has been utilized in the context of
calculating strong interaction corrections to various processes.
Recently, the application of effective field theories in heavy quark
systems has led to great progress in our understanding of weak
decays. In particular, heavy quark effective field theory (HQET) has
been utilized to study hadrons composed of one heavy quark. The use of
HQET allows us to separate the physics stemming from the two scales
which are relevant to heavy-light bound states, namely the heavy mass
$m$ and the strong interaction scale $\Lambda_{QCD}$.  In a seminal
paper\cite{cl} Caswell and Lepage introduced a similar effective field theory
to study non-relativistic bound states.  However, this theory differs
from HQET in several very important ways. The description of 
non-relativistic bound
states is  complicated by existence of the small parameter 
$v\sim\alpha(mv)$ in the effective theory\footnote{This complication is
often said to lead to the problem of having many scales in the theory
$m,~mv,~ mv^2\ldots$}, where $v$ is the relative velocity of the
particles of mass $m$ which compose the bound state. Furthermore, in
HQET the heavy quarks are labeled by velocities which are unchanged by
bound state effects at leading order. Velocity changing weak
transitions are accounted for by ``integrating in'' quarks of varying
velocities \cite{hg}.  In heavy-heavy systems it is no longer true
that the quark\footnote{We will refer to the bound state constituents
as quarks though they may be electrons as well. Furthermore, we call
the gauge particles gluons to generalize to the Non-Abelian case.}  
velocity is
fixed. Indeed Coulombic, velocity altering exchanges are what builds
up the Schr\"{o}edinger kernel.  Thus, it is clear that the effective
Lagrangian for the two systems should be dissimilar. Indeed, in HQET an
operator scales in $1/M$ according to its dimension, whereas as in
non-relativistic gauge theories this is not so unless one modifies
the Lagrangian, as will be shown below.

HQET relies upon an expansion in the heavy quark mass, whereas
Non-relativistic Gauge Effective Field Theories (NRGT) are expansions
in the relative velocity, or equivalently, as we shall couch it $c$,
the speed of light.  Though it is well known that certain identical
operators of the same mass dimension may be of different orders in
their respective expansion (most notably $\psi^\dagger ({\bf
D}^2/2m)\psi)$, it has not been pointed out that even for fixed
dimensions the operators of the two theories will in general be
different. For instance, if one uses dimensionally regulates HQET
mixing can only occur between operators with the same scalings
in $1/M$, this will not be true in non-relativistic field
theories unless one modifies the Lagrangian, as will be shown
in this letter.

In the effective field theory formalism we write down a low energy
Lagrangian which reproduces the S matrix elements (in some cases the
Greens functions) of the full theory up to some chosen order in a
double expansion in the coupling $\alpha$ and some other parameter
which dictates the size of the matrix elements in the low energy
theory, the heavy quark mass in the case of HQET.  The difference
between the full theory and the effective theory lies in the
ultraviolet modes. This difference is accounted for in the 
low energy effective 
theory by the proper choice of coefficients in the Lagrangian (the
``matching'' procedure). The utility of the effective theory lies in
the fact that calculations in the effective theory are much simpler
now that all the short distance physics has been trivialized.

A crucial part of the matching procedure is the bookkeeping of the
expansion parameters. For instance, in HQET where the expansion
parameters are $\alpha_s$ and $1/m$, it is possible to place all the
dependence on the expansion parameters into overall coefficients of
operators in the Lagrangian.  This is helpful for two reasons: In
performing the matching, operators which are formally of higher order
do not contribute to the coefficients of lower order operators.  Second,
the order at which matrix elements of operators enter should be
determined by the coefficients of the operators.  That is, the states
should not depend on the expansion parameters in such a way that the
power counting is jeopardized.  
For example, in HQET the states are
independent of $m$ since the lowest order Lagrangian is independent of
$m$ and the normalization of the states is chosen to be
\beq
\langle{\vec v}\,^\prime\mid \vec v\,\rangle 
= 2v^0(2\pi)^3\delta^3(\vec{v^\prime}-\vec v)
\eeq

For the case of NRGT's the expansion parameters are $\alpha$ and $v$,
the relative quark velocity. One immediately sees that things will be
more difficult in this case since the expansion parameter $v$ is
dimensionless. This complication led to velocity scaling rules,
derived via heuristic arguments, which assigned powers of $v$ to
fields, operators and derivatives \cite{vex}.  A simpler bookkeeping
method was presented in \cite{ml} where the authors rescale fields and
coordinates by $v$ in such a way as to make explicit the powers of $v$
in the Lagrangian. Here we introduce a slightly different approach
which follows simply via an expansion in the now dimensionful
parameter $1/c$.  To the extent that the physical system is truly
non-relativistic, operators with velocity dimension $n$ are of
magnitude $\sim(v/c)^n$, with $v$ a {\em dynamically} generated
scale. This is analogous to HQET where operators of mass dimension $n$
are of magnitude $(\Lambda_{\rm QCD}/m)^n$. Moreover the $1/c$
expansion forces one to modify the Lagrangian.

Let us consider the large $c$ limit of a non-Abelian gauge theory.
In this limit the Lagrangian density is given by\cite{qed},
\beqa
\label{cexp}
{\cal L}&= &
\frac{1}{2}(\partial_iA^a_0-\frac{1}{c}\partial_0A_i^a-
\frac{g}{c}f_{abc}A^b_iA^c_0)^2
-\frac{1}{4}(\partial_i A^{a}_j-\partial_jA^a_i-\frac{g}{c}f_{abc}A^b_i
A^c_j)^2 \nonumber \\
&+&
\psi^\dagger\left(iD_0 +\frac{{\bf {D}}^2}{2m}\right)\psi 
+\frac{c_F}{2mc}
\psi^\dagger\bfsigma  \cdot {\bf B}\psi
+\frac{1}{8m^3c^2}\psi^\dagger{\bf {D}}^4\psi  \nonumber \\
&+&\frac{c_D g}{8m^2c^2} \psi^\dagger 
({\bf D}\cdot{\bf E}-{\bf E}\cdot{\bf D})\psi + \frac{c_S}{8m^2c^2}\psi^\dagger\bfsigma  \cdot({\bf {D}}\times {\bf{E}}-
{\bf {E}}\times {\bf{D}})\psi+O(1/c^3),
\eeqa 
where 
\beq
D_0=\frac{\partial}{\partial t}-gA_0; ~~{\bf {D}}=
{\bf \nabla}-\frac{g}{c}{\bf {A}},
\eeq
and $\psi$ is
a non-relativistic  2-spinor describing the heavy quark. In
 addition we have rescaled
the fermion field by a factor of $\sqrt c$. 
For simplicity we have omitted a 2-spinor describing the heavy
antiquark.
The constants $c_F$, $c_D$ and $c_S$
are determined by matching onto the full theory.

The explicit powers of the dimensionful
parameter $1/c$ in (\ref{cexp}) now makes the power counting simple. 
The lowest order Lagrangian is
\beq
\label{cexpzeroth}
{\cal L}_0= 
\frac{1}{2}(\partial_iA^a_0-\frac{1}{c}\partial_0A_i^a)^2
-\frac{1}{4}(\partial_i A^{a}_j-\partial_jA^a_i)^2
+\psi^\dagger\left(iD_0 +\frac{{\bf \nabla}^2}{2m}\right)\psi.
\eeq 
Notice that we have retained the $1/c$ term
in the kinetic energy of the transverse gluons in the lowest
order Lagrangian. That this is necessary is easily seen 
by considering the Hamiltonian, in which the
coefficient of the kinetic energy is $c^2$. 
The eigenstates of
this lowest order Hamiltonian constitute the states of the effective
theory.

At this point we still have not accomplished what we set out to do, 
namely, trivialize the $c$ dependence of the Lagrangian. Eq. 
(\ref{cexpzeroth}), as it stands, will lead to a transverse gluon propagator with non-trivial
$c$ dependence which can, and as we shall see below does, jeopardize
the power counting in $1/c$. To  fix this problem we have a choice 
to either rescale the time or spatial coordinates of the gauge field by $c$.
However, rescaling the time coordinate of the gauge field is unacceptable 
as it will
destroy the initial value problem because,  the Hamiltonian for the
gauge field and fermion fields would then  depend on different time 
coordinates. Thus, we make the rescaling
\beq
\tilde A_i(\vec y=\vec x/c,t) = \sqrt c A_i(\vec x,t),
\eeq
leaving the Coulomb gauge Lagrangian
\beqa
\label{rescale}
L_0=\int d^3y
\frac{1}{2}[(\partial_0\tilde A_i^a)^2
-(\partial_i \tilde A^a_j)^2]
+
\int d^3x\left[
\psi^\dagger  \left(iD_0 +\frac{\bfpartial^2}{2m}\right)\psi+
\frac{1}{2}(\partial_iA_0^a)^2\right]~
.
\eeqa 
We choose to work in the Coulomb gauge for the rest of the paper since
it is the most natural choice in a non-relativistic theory. Moreover,
it allows for the clean separation of powers of $1/c$ as is clear
from (\ref{rescale}).
Note that the states of the effective theory are the eigenstates of the lowest
order Hamiltonian derived from this Lagrangian and, as such,  are unconfined
Coulombic bound states.  While these states are independent of $c$
they are not independent of $g$. The Coulomb gluons are leading order
and are not treated perturbatively.  We will return to the issue of
confining effects at the end of the paper.

Now if we follow through with the consequences of the $1/c$ expansion we will
be forced to incorporate the multipole expansion. To see this, let us
now consider the $1/c$ corrections in (\ref{cexp}), concentrating for the
moment on the Abelian pieces (the extension to the non-Abelian case follows
trivially). The leading $1/c$ corrected fermionic bilinear 
Lagrangian is
given by
\beq
\label{leaddip}
L_{c^{-1}}=\int d^3x \psi^\dagger(t,\vec{x})\left[
\frac{e}{mc^{3/2}}\tilde A_i(t,
\vec{x}/c)\frac{\partial}{\partial x^i}+
\frac{c_F}{2mc^{5/2}}\bfsigma \cdot {\bf \tilde B}(t,\vec{x}/c)\right]
\psi (t,\vec{x}).
\eeq
Expanding in $1/c$ leads to the multipole expansion
\beqa
\label{dip}
{\cal L}_{\rm mp}=\frac{e}{mc^{3/2}}\psi^\dagger
(t,\vec{x})&&\!\!\!\!\!\!\!\!\!
\!\!\!\left({\bf
\tilde A}(t,\vec0) +\frac{x_i}{c}
\nabla\cdot{\bf \tilde  A}(t,{\vec0}) 
\frac{\bfpartial}{\partial x^i}+\cdots\right) \psi (t,\vec{x})  \nonumber \\
&+&
\frac{c_F}{2mc^{5/2}}\psi^\dagger(t,\vec{x})\bfsigma \cdot {\bf \tilde B}
(t,\vec0) \psi (t,\vec{x})+\cdots~,
\eeqa
where
\beq
  {\bf \tilde B}=\epsilon_{ijk}\frac{\partial}{\partial (y^j)}\tilde 
A_k(t,
\vec{y}).
\eeq
The expansion breaks translational as well as gauge invariance which are
symmetries that are restored at each order in $1/c$, and the coefficients
in front of each operator are therefore fixed.

Now let us consider the matching procedure for this
theory. This will elucidate the power counting scheme as well
as the problems one runs into if the multipole expansion is
not performed.
Using standard diagrammatic techniques it is easy to show that in the full
theory the one particle irreducible amputated $N$ point functions with 
$L$ loops has an overall factor of
 \footnote{Since $\alpha =e^2/c$ we could equally
write the expansion purely in terms of $1/c$ taking
$e\approx 1$. We have chosen to write the expansion in terms of
$\alpha$ in analogy with the expansion in HQET.}
\beq \Gamma^{N}\propto (\alpha)^{L-1+N/2}c^{-N/2}.
\eeq
Use of dimensional regularization leads to the functional dependence
$\Gamma^{N}(q_i\cdot q_j/(mc)^2,\mu^2/(mc)^2,\epsilon)$.  On shell
these corrections will be both IR as well as UV divergent.  The UV
divergences are taken care of by renormalizing the full theory, while
the IR divergences will cancel with those in the effective theory,
since both theories behave the same in the infrared. To match onto the
effective theory we then expand this full theory result in a power
series in $1/c$. The coefficients $c_i$ are then chosen so that the
above expansion is reproduced by the effective theory, which we now
discuss.

At tree level the matching is trivial. Beyond the terms in (7) the
effective theory Hamiltonian will also contain spatially non-local,
instantaneous four quark operators which scale as $1/c^2$. These
operators arise as a consequence of taking the leading term in the $c$
expansion of full theory diagrams with transverse gluon exchange
between quarks.

Let us now study the one loop correction
in the effective theory. We will first perform the  calculation
using (\ref{cexp}) and show that operators which are supposedly of higher
order in $1/c$ will generate  lower order operators, even within
dimensional regularization. 
We will then show
that using (\ref{leaddip}) no such mixing occurs, even within dimensional
regularization.
Let us consider the calculation of the correction to the two point function coming from two insertions of the magnetic moment operator, $\sigma
\cdot {\bf B}$. If one wishes to keep the power counting such that
a given operator should scale as a fixed power in $1/c$ (as dictated
by its overall coefficient), 
the higher order operators will not  contribute to the renormalization
of the lowest order Lagrangian. 
That this is so in heavy quark effective
theory is easy to see simply on dimensional grounds.

Using the Feynman rules derived from (\ref{cexp}) we have
\beq
\label{oneloop}
i\Gamma^{(2)}=\frac{c_F^2 C(R) e^2}{4m^2c^2}\int\! \frac{d^n k}
{(2\pi)^n}\;\frac{(\vec{k}\times \vec{\sigma})\cdot (\vec{k}
\times \vec{\sigma})
}{(k_0^2/c^2-\vec{k}^2+i\epsilon)
(E +k_0-(\vec{p}+\vec{k})^2/2m+i\epsilon)}.
\eeq
We integrate over $k_0$ and choose to close the contour in the
upper half plane, picking up only the negative energy pole.
Integrating over the magnitude of $\vec{k}$ leaves
\beq
 \Gamma^{(2)}=\frac{c_F^2C(R)\alpha}{\pi} \Gamma(2\epsilon)\left[
2(E-
\frac{\vec{p}\;^2}{2m})+4mc^2+\frac{4}{3}\frac{\vec{p}\;^2}{2m}
\right]+{\rm finite},
\eeq
where $C(R)$ is $1$ and $4/3$ in QED and QCD, respectively.
We see that using (\ref{cexp}) leads to the mixing of operators of
different orders in $1/c$.
This can be avoided by expanding (\ref{oneloop}) in powers of $\vec{k}$. 
In previous matching calculations \cite{kn,l,m} this 
is in fact  what has been done, and it is 
justified as a bona-fide approximation,
as an expansion in small $\vec{k}$\cite{kn}. 
This amounts to dropping the $\vec{k}$ dependent terms in
the denominators. 
Such an expansion makes working with dimensional
regularization particularly simple, since we now have a scaleless
integral which vanishes in this scheme. 

However, we emphasize that, 
once we choose to amend the Feynman rules we must
also amend the low energy theory. Not to do so would destroy the power
counting scheme. In Ref.~\cite{ml}, the authors point out that
the transverse gluons can lead to enhancements in $c$ (their $1/v$) in the low
energy theory. 
This is only true if one insists upon calculating the
low energy matrix elements using (\ref{cexp}). Instead one must calculate
using an amended Lagrangian (8) which reproduces the Feynman rules utilized in
the matching calculations.
Indeed it is simple to show that Feynman rules in  (\ref{dip})
leads to the necessary expansion.  

Using (\ref{dip}) we may now calculate anew the contribution to
the two point function from two insertions of $\sigma\cdot B$,
\beq
i\Gamma^{(2)}=\frac{c_F^2 C(R)e^2}{2m^2c^5} \int \frac{d^nk}{(2\pi)^n}
\frac{\vec{k}^2}{(k_0^2-\vec{k}^2+i\epsilon)(E-\vec{p}\;^2/2m+k_0+
i\epsilon)}.
\eeq
Notice that all the $c$ dependence is now explicit, and the variable
$\vec{k}$ has units of energy.
A simple calculation leads to the result
\beq
\Gamma^{(2)}=\frac{c_F^2C(R)\alpha}{2\pi}\Gamma(2\epsilon)\frac{
(E-\frac{\vec{p}^2}{2m})^3}{m^2c^4}+{\rm finite}.
\eeq
This correction renormalizes some higher order operator which
vanishes by the equations of motion. It is clear that once
we use the correct effective, theory the corrections resulting from
the insertion of higher order operators will never feed down
into the matching for lower order operators.

Just as the $1/c$ expansion dictates the proper matching procedure, it
also greatly simplifies the power counting rules in the
low energy theory. In HQET the magnitude of a matrix element is dictated
by the explicit power of $1/m$ in its coefficient. Thus, an operator
of mass dimension $d$ with a factor of $m^{-n}$ in its coefficient 
is of order $\frac{\Lambda_{QCD}^{n+4-d}}{m^n}$. In NRGT's, as formulated
here, the power counting is just as simple. All the powers of
$1/c$ have been made explicit, and we
can read off the size of a matrix element simply by counting powers
of $c$ and doing dimensional analysis.

As an example of the power counting procedure let us consider the pedagogical 
example of
the relativistic corrections to the bound state
energy of Hydrogen\cite{lz}. Given that 
$\alpha$ and $v/c$ are of the same order, we must calculate the
matching corrections to the appropriate order in $\alpha$ for the
accuracy we wish to attain. 
The leading corrections come from pure $(v/c)^2$ corrections stemming
from the effective Lagrangian.  The first such term is the correction
to the kinetic energy $\psi^\dagger\bfpartial^4/(8m^3c^2)\psi$. By
dimensional analysis, its matrix element is 
$mv^2(v/c)^2$, which yields a relative contribution of order
$(v/c)^2$.  There are no $\alpha$ corrections in the matching to this
operator.  Next there are $(v/c)^2$ corrections coming from the Darwin
term and the spin orbit coupling which are again of relative order
$(v/c)^2$, since we may pick out the Coulombic piece of the
electric field. Both the Darwin term and the spin orbit term will get
matching corrections at order $\alpha$ and will thus give a
contribution to the energy shift at relative order $(v/c)^3$ as well.  There
are no further corrections at order $(v/c)^2$, assuming the proton to be
infinitely heavy so that the magnetic interactions become
irrelevant. There is a correction of relative order $(v/c)^3$ coming from
corrections to the Coulomb potential due to pair creation which are
accounted for in the matching and lead to a term in the effective Lagrangian
given by
\beq
O_U=c_U \alpha \frac{\partial_iA_0\partial_j^2\partial_iA_0}{m^2c^2}.
\eeq
Finally, we come to the energy shift due to transverse photon
propagation in the bound state. This is a self-energy correction of the
electron propagating in the Coulomb background and is of relative
size $(v/c)^3$ due to a factor of $c^{-3/2}$ coming from each
transverse photon vertex. This 
agrees with the well known result for the Lamb shift.

Let us now return to the issue of confinement in the Non-Abelian theory.
While it may be surprising that the non-Abelian couplings are
subleading, it is clear that the non-Abelian nature of the theory
should be irrelevant to the details of the bound state as its size is
reduced. The confining effects in a Coulombic bound state should be
suppressed by powers of $\Lambda_{QCD}/m$. Setting $r\sim 1/mv$ in the
virial theorem
\beq
\label{vir}
\frac{\alpha_s(1/r)}{r}\sim \frac{mv^2}{c},
\eeq
gives $v/c\sim\alpha_s(m v)$, which leads to the conclusion that, at
small $v$, $\Lambda_{QCD}/m$ scales like\footnote{We will take  
$\Lambda_{QCD}$ to have the units of mass.}
 $v/c$.
The question then becomes how do we properly take into account the
effects of confinement in this effective field theory?
Since confinement will not arise in perturbation theory we must
insist that the zeroth order states contain the confining potential.
The omission of the non-perturbative effects will lead to the breakdown
of the $1/c$ expansion as we will now show.

If we include the full non-Abelian gluondynamics in the zeroth order 
effective theory, then 
the Coulombic potential will now be modified by the
linear rise due to confinement. 
At first this may seem a bit worrisome since the non-Abelian
piece has explicit factors of $1/c$ in front of them and thus the states
of the theory will depend upon $c$, which is exactly what we were
trying to avoid. However, this dependence on $c$ will not destroy the
systematics.
The analytic dependence now  implicit in the matrix elements will
 clearly not upset  the systematics of the $1/c$ expansion.
The non-analytic dependence on $c$ will be introduced through factors
such as 
\beq
e^{\kappa c/g^2}=\left( \frac {\Lambda_{QCD}}{m}\right)^P,
\eeq
where $P$ is some positive power. Thus, the $c$ dependence of the
matrix elements due to non-perturbative effects 
can only introduce higher order corrections.  If we chose to use the
Coulombic states, we would never see these higher order effects, and  
since $\Lambda_{QCD}/m \sim v/c$, we would  possibly miss effects
of the same order we wish to keep.

We have shown that Non-Relativistic Gauge Theory (NRGT) is
conveniently organized in temrs of 
 an expansion in the dimensionful parameter
$1/c$. Furthermore, if dimensional regularization is used, the
organization of the expansion is very simple, there is no operator
mixing across different orders in the expansion and the order of the
operators in the expansion can be read off directly from their dimensions 
without resort to heuristic arguments. We emphasize, however, that dimensional regularization is not
mandatory. 
Indeed one may formulate NRGT's with, for example, a momentum cut-off
$\Lambda$. If $\Lambda\ll m$, with $m$ the quark mass, then the
multipole expansion is automatic.  However, one then has a triple
expansion, in $\alpha$, $1/c$ and $\Lambda/m$.  It is not necesary to
choose $\Lambda<m$; with $\Lambda\gg m$ one can work with a double
expansion, in $\alpha$ and $1/c$ only.\footnote{In fact, it is
unnecesary to perform a multipole expansion, if one is willing to fine
tune the coefficients in the expansion order by order in perturbation
theory.} In either case operators of different orders in the $1/c$
expansion mix. The organization by $1/c$ is still useful because once
the matching procedure has been completed the order at which an
operator enters can still be trivially obtained from its velocity
dimension.

\vskip1in
\noindent{\bf Acknowledgments}

We thank Mark Wise for many discussions and for
 emphasizing that we should explore the full
consequences of the $1/c$ expansion. We also M. Luke, A. Falk and 
S. Sharpe for helpful comments. This work was
supported in part by the Department of Energy under Grant 
No.~DOE-FG03-90ER40546.

\end{document}